\documentclass[aps,prl,twocolumn,amsmath,amssymb,floatfix,reprint,nofootinbib,superscriptaddress,longbibliography]{revtex4-2}
\usepackage{graphicx}
\usepackage{amsmath}
\usepackage{amssymb}
\usepackage{braket}
\usepackage{booktabs}
\usepackage{dcolumn}
\usepackage{bm}
\usepackage{mathrsfs}
\usepackage{dsfont}
\usepackage{textcomp}
\usepackage{multirow}
\usepackage{xcolor}
\usepackage[colorlinks=true,linkcolor=blue,citecolor=red,urlcolor=blue]{hyperref}
\usepackage{url}
\begin{document}

\title{Qubit Readout via State-Dependent Radiative Linewidths}

\author{Yiming Yu}
\affiliation{Fujian Key Laboratory of Quantum Information and Quantum Optics, College of Physics and Information Engineering, Fuzhou University, Fuzhou 350116, China}

\author{Xinyu Zhao}
\affiliation{Fujian Key Laboratory of Quantum Information and Quantum Optics, College of Physics and Information Engineering, Fuzhou University, Fuzhou 350116, China} 

\author{Jia-Wen Yu}
\affiliation{Fujian Key Laboratory of Quantum Information and Quantum Optics, College of Physics and Information Engineering, Fuzhou University, Fuzhou 350116, China}

\author{Ke-Xiong Yan}
\affiliation{Fujian Key Laboratory of Quantum Information and Quantum Optics, College of Physics and Information Engineering, Fuzhou University, Fuzhou 350116, China}

\author{Wei~Qin}
\affiliation{Center for Joint Quantum Studies and Department of Physics, School of Science, Tianjin University, Tianjin 300350, China}

\author{Ye-Hong Chen}
\email{yehong.chen@fzu.edu.cn}
\affiliation{Fujian Key Laboratory of Quantum Information and Quantum Optics, College of Physics and Information Engineering, Fuzhou University, Fuzhou 350116, China}
\affiliation{Quantum Information Physics Theory Research Team, Center for Quantum Computing, RIKEN, Wako-shi, Saitama 351-0198, Japan}

\author{Yan Xia}
\email{xia-208@163.com}
\affiliation{Fujian Key Laboratory of Quantum Information and Quantum Optics, College of Physics and Information Engineering, Fuzhou University, Fuzhou 350116, China}

\author{Franco Nori}
\affiliation{Quantum Information Physics Theory Research Team, Center for Quantum Computing, RIKEN, Wako-shi, Saitama 351-0198, Japan}
\affiliation{Department of Physics, University of Michigan, Ann Arbor, Michigan 48109-1040, USA}

\date{\today}

\begin{abstract}

Fast qubit readout conventionally encodes state information in a dispersive frequency shift. 
Here we formulate a linewidth-encoded quantum non-demolition measurement channel in which the qubit state enters the external radiative amplitude, equivalently a state-dependent Lindblad jump operator.
Starting from an empty cavity, we show analytically that this dissipative channel imprints state information on the output field at $O(t)$, whereas standard dispersive readout starts at $O(t^2)$ because it requires intracavity buildup and conditional phase accumulation. 
This short-time scaling produces faster matched-filter signal-to-noise ratio accumulation and persists in finite-resource comparisons, including photon-number limits, external-linewidth budgets, cavity depletion, and pulse-optimized dispersive baselines.
We further outline an auxiliary-mode route that converts a qubit-state-dependent auxiliary susceptibility into a state-dependent linewidth.
These results identify engineered dissipation as an information-carrying resource for fast quantum non-demolition readout.

\end{abstract}

\maketitle
\begingroup
\renewcommand{\thefootnote}{\fnsymbol{footnote}}
\footnotetext[1]{yehong.chen@fzu.edu.cn}
\footnotetext[2]{xia-208@163.com}
\endgroup

\textit{Introduction.}---Fast qubit readout is a central primitive for quantum information processing. 
quantum non-demolition (QND) measurement theory describes how information can be extracted while preserving the measured logical basis~\cite{ref1,BraginskyKhalili1996,ref2}, while Markovian master equations and input-output theory connect the intracavity dynamics to the measured microwave field~\cite{GoriniKossakowskiSudarshan1976,ref3,BreuerPetruccione2002,LidarOpenQuantumSystems2019,ref5,ref6}. 
In superconducting circuit quantum electrodynamics (QED), the standard implementation is dispersive readout: the qubit shifts the resonator frequency, and the outgoing microwave phase carries the state information~\cite{YouNoriPRB2003,YouNoriPhysicsToday2005,GuMicrowavePhotonics2017,KockumNoriJosephsonQubits2019,KrantzQuantumEngineer2019,KjaergaardCurrentState2020}. 
This mechanism has enabled high-fidelity single-shot readout~\cite{ref11,ref12,ref13,ref14,ref15,ref20,DassonnevilleCrossKerr2020,PhysRevLett.133.233605,GardTunableReadout2024,StefanskiFluxPulse2024,ChappleBalancedCrossKerr2025}, quantum trajectories~\cite{ref16,ref17}, and measurement-based feedback or parity measurements~\cite{ref18,ref19,KellyNature2015,HinderlingParityReadout2024}.

Despite these successes, standard dispersive readout has a transient bottleneck when starting from an empty resonator.
The cavity field must first ring up before the qubit-induced frequency shift can generate conditional phase separation~\cite{Touzard2019}.
Thus, under photon-number and finite-time constraints, the output distinguishability appears only after this two-step buildup. 
In practice, increasing the drive power or measurement bandwidth is constrained by cavity nonlinearity, measurement-induced transitions, and Purcell-like relaxation channels~\cite{ref21,ref22,ref23,CatelaniFlux2011,CatelaniFrequencyShift2011,WangIdealQND2019,CatelaniDecoherence2012,SunadaIntrinsicPurcell2022}, motivating Purcell-filter engineering~\cite{ref26,ref27,ref28,GoviaClerkSqueezing2017,KonoJQF2020}.
Longitudinal-coupling readout provides another fast-QND route by using a qubit-state-dependent Hamiltonian force to displace the pointer~\cite{ref29,ref30,ref31,IkonenMultichannel2019,BottcherLongitudinal2022,HarptLongitudinalReadout2025,XuCatLongitudinal2026}. 
In continuous microwave readout, however, the measured information is carried away by an output channel, whose coupling operator need not be a passive element.
Reservoir engineering shows that Lindblad jump operators can be designed as physical resources~\cite{ref32,ref33,ref34,ref35,ref36}, and state-dependent cavity decay has been observed in bath-mediated circuit-QED systems~\cite{WangNonreciprocity2024}.

In this manuscript, we formulate and analyze a linewidth-encoded dissipative readout channel in which state information enters through the radiative coupling to the output port, rather than through a Hamiltonian frequency shift or force.
In the present protocol, the qubit state is encoded in the external linewidth, equivalently in the coefficient of a state-dependent Lindblad jump operator.
Because the jump operator is diagonal in the measured $\sigma_z$ basis, the ideal channel is QND at the population level.

This change of measurement generator has an immediate dynamical consequence.
Starting from an initially empty resonator, we show analytically that the output-field separation starts at $O(t)$ for the dissipative channel but only at $O(t^2)$ for standard dispersive readout, which must first build intracavity phase separation.
This produces faster matched-filter signal-to-noise ratio (SNR) buildup over a finite short-time window.
We then test the mechanism under finite photon-number, external-linewidth, and pulse-control resources, and outline an auxiliary-mode route in which a QND dispersive shift of a lossy auxiliary mode is converted into a state-dependent linewidth.
Together, these results identify engineered dissipation as an information-carrying resource for fast QND readout.

\textit{Measurement dynamics and short-time information flow.}---Let $P_s=|s\rangle\langle s|$ with $s\in\{e,g\}$ denote the projectors onto the measured qubit basis, and let $\sigma_z|s\rangle=z_s|s\rangle$ with $z_e=+1$ and $z_g=-1$. At the Markov input-output level, the fixed-qubit-state dynamics takes the standard single-port form~\cite{ref5},
\begin{gather}
\dot a_s(t)
=
-\left(i\Delta_s+\frac{\kappa_s}{2}\right)a_s(t)
-\sqrt{\kappa_s}\, b_{\rm in}(t),
\label{eq:dissipative-langevin}
\\[-0.2em]
b_{\rm out}^{(s)}(t)
=
b_{\rm in}(t)
+
\sqrt{\kappa_s}\, a_s(t).
\label{eq:dissipative-input-output}
\end{gather}
Here $a_s(t)$ denotes the Heisenberg-picture readout-mode annihilation operator conditioned on a fixed qubit state $s$. The fields $b_{\rm in}(t)$ and $b_{\rm out}^{(s)}(t)$ are the input and output fields in the measurement line, $\sqrt{\kappa_s}$ is the state-dependent radiative amplitude, $\kappa_s$ is the corresponding external linewidth, and \(\Delta_s\) is the state-dependent effective detuning.
In this ideal model, we choose the port phase convention in which the radiative amplitudes are real and positive.

\begin{figure}[t]
    \centering
    \includegraphics[width=0.45\textwidth]{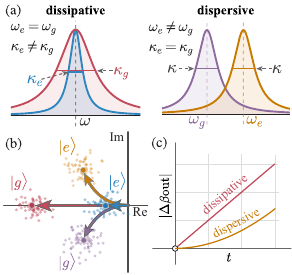}
    \caption{
        Conceptual comparison of linewidth-encoded dissipative readout and conventional dispersive readout.
        (a) Spectral response: the dissipative channel keeps the readout frequency state independent, $\omega_e=\omega_g=\omega$, but makes the external linewidth state dependent, $\kappa_e\neq\kappa_g$; the dispersive channel instead uses $\omega_e\neq\omega_g$ with $\kappa_e=\kappa_g=\kappa$.
        (b) Schematic output-pointer distributions in the complex output-amplitude plane. Red/blue denote the $|g\rangle/|e\rangle$ dissipative responses; purple/orange denote the $|g\rangle/|e\rangle$ dispersive responses.
        (c) Leading short-time output separation from an initially empty resonator.
        }        
    \label{fig:1}
\end{figure}

In the model above, the resonator response is governed by two state-resolved parameters: the effective detuning \(\Delta_s\) and linewidth \(\kappa_s\). 
We isolate the linewidth-encoded limit by imposing:
$\kappa_e\neq\kappa_g,~\Delta_e=\Delta_g\equiv\Delta.$
Thus the qubit state is encoded in the external linewidth rather than in a resonator-frequency shift.
Figure~\ref{fig:1} summarizes the resulting contrast with conventional dispersive readout, from the spectral response to the output-pointer geometry.

For a fixed qubit state, Eq.~\eqref{eq:dissipative-input-output} identifies the port coupling operator as $L_s=\sqrt{\kappa_s}\,a$. 
Promoting the fixed-state amplitude $\sqrt{\kappa_s}$ back to a diagonal operator on the qubit Hilbert space gives the effective Lindblad jump operator
\begin{equation}
L
=
\left(
\sqrt{\kappa_e}\, P_e+\sqrt{\kappa_g}\, P_g
\right)a.
\label{eq:lindblad-jump}
\end{equation}
Under the Markov approximation, the reduced system obeys
$\dot\rho=-i[H,\rho]+\mathcal D[L]\rho$. 
Within this Markovian Lindblad dynamics~\cite{BreuerPetruccione2002,LidarOpenQuantumSystems2019,ref3}, the QND condition is expressed as a Heisenberg-picture conservation law for the measured observable~\cite{BraginskyKhalili1996}: $d\sigma_z/dt=i[H,\sigma_z]+\mathcal D^\dagger[L](\sigma_z)$.
In the ideal model, the Hamiltonian contains no transverse qubit term in the measured basis, so $[H,\sigma_z]=0$. Since $L$ is built from projectors in the $\sigma_z$ basis and the readout-mode operator $a$, Eq.~\eqref{eq:lindblad-jump} gives $[L,\sigma_z]=0$ and hence $[L^\dagger L,\sigma_z]=0$. The dissipative contribution therefore vanishes, $\mathcal D^\dagger[L](\sigma_z)=L^\dagger\sigma_zL-\{L^\dagger L,\sigma_z\}/2=0$. 
Thus $d\sigma_z/dt=0$ at the level of the effective master equation, so the measurement is QND with respect to the $\sigma_z$ basis.

For a constant coherent input drive, $\langle b_{\rm in}(t)\rangle=\beta$, we denote the conditional mean readout-mode amplitude by $\alpha_s(t)=\langle a_s(t)\rangle$ and take the initial condition $\alpha_s(0)=0$.
Solving Eq.~\eqref{eq:dissipative-langevin} for the coherent input drive gives
\begin{equation}
\alpha_s(t)
=
-
\frac{\sqrt{\kappa_s}\,\beta}
{i\Delta+\kappa_s/2}
\left[
1-
e^{-(i\Delta+\kappa_s/2)t}
\right].
\label{eq:dissipative-alpha}
\end{equation}
The corresponding output amplitude $\beta_{\rm out}^{(s)}(t)=\langle b_{\rm out}^{(s)}(t)\rangle$ is
\begin{equation}
\beta_{\rm out}^{(s)}(t)
=
\beta
-
\frac{\kappa_s\beta}
{i\Delta+\kappa_s/2}
\left[
1-
e^{-(i\Delta+\kappa_s/2)t}
\right].
\label{eq:dissipative-output-kappa}
\end{equation}
Expanding around $t=0$ gives
\begin{equation}
\beta_{\rm out}^{(s)}(t)
=
\beta
-
\kappa_s\beta t
+
O(t^2).
\label{eq:dissipative-output-short}
\end{equation}
Therefore the output-field difference between the two qubit states is
\begin{equation}
\Delta\beta_\kappa(t)
\equiv
\beta_{\rm out}^{(e)}(t)
-
\beta_{\rm out}^{(g)}(t)
=
-
(\kappa_e-\kappa_g)\beta t
+
O(t^2).
\label{eq:dissipative-delta-beta-short}
\end{equation}
Thus linewidth-encoded readout generates output-field separation already at first order in time.

For comparison, the conditional dynamics of standard dispersive readout is~\cite{ref9}
\begin{gather}
\dot a_s(t)
=
-\left[
i(\Delta+z_s\chi)+\frac{\kappa}{2}
\right]a_s(t)
-\sqrt{\kappa}\,b_{\rm in}(t),
\label{eq:dispersive-langevin}
\\[-0.2em]
b_{\rm out}^{(s)}(t)
=
b_{\rm in}(t)
+
\sqrt{\kappa}\,a_s(t),
\label{eq:dispersive-input-output}
\end{gather}
where $\kappa$ is the state-independent external linewidth and $\chi$ is the dispersive half shift. The two qubit states therefore have the same external linewidth $\kappa$ but different detunings, $\Delta_s=\Delta+z_s\chi$. 
In contrast, for dispersive readout the first-order output response is state independent. We therefore expand one order further to expose the leading state-dependent contribution:
\begin{equation}
\beta_{\rm out}^{(s)}(t)
=
\beta
-
\kappa\beta t
+
\frac{\kappa\beta}{2}
\left[
i(\Delta+z_s\chi)
+
\frac{\kappa}{2}
\right]t^2
+
O(t^3).
\label{eq:dispersive-output-short}
\end{equation}
Equations~\eqref{eq:dissipative-output-short} and \eqref{eq:dispersive-output-short} show the different output-separation mechanisms illustrated in Fig.~\ref{fig:1}(b). 
Thus the linewidth-encoded channel separates the output pointers at $O(t)$ along the amplitude quadrature, whereas dispersive readout has a common $O(t)$ displacement and acquires state-dependent phase-quadrature separation only at $O(t^2)$.
Taking the difference between the two dispersive output fields gives
\begin{equation}
\Delta\beta_\chi(t)
=
i\kappa\chi\beta t^2
+
O(t^3).
\label{eq:dispersive-delta-beta-short}
\end{equation}
These results give the two separation orders indicated in Fig.~\ref{fig:1}(c):
$|\Delta\beta_\kappa|\propto t$ and $|\Delta\beta_\chi|\propto t^2$.
We quantify the accumulated distinguishability using the matched-filter signal-to-noise ratio (SNR)~\cite{ref4},
\begin{equation}
{\rm SNR}^2(t_f)
=
\frac{1}{S_{\rm eff}}
\int_0^{t_f}
\left|
\Delta\beta_{\rm out}(t)
\right|^2dt,
\label{eq:matched-filter-snr}
\end{equation}
where $S_{\rm eff}$ is the effective output-noise spectral density, with the same normalization used in the Gaussian-overlap error estimate below.
Substituting the leading short-time terms yields
\begin{equation}
    {\rm SNR}_\kappa
    \propto
    |\kappa_e-\kappa_g|
    |\beta|
    t_f^{3/2},~~{\rm SNR}_\chi
    \propto
    \kappa|\chi|
    |\beta|
    t_f^{5/2}.
    \label{eq:dissipative-snr-scaling}
\end{equation}

The dissipative channel therefore builds SNR faster in the short-time regime. 
A non-asymptotic finite-window bound for the constant-input model is given in the Supplemental Material~\cite{SupplementalMaterial}.

\textit{Finite-resource performance.}---We now move from the analytic scaling to finite-resource readout performance. 
The comparison is organized around finite-time readout under experimentally relevant budgets: the intracavity photon number \(N_{\max}\), the readout duration \(t_f\), the external-linewidth scale and contrast \((\kappa_{\max},r_\kappa)\), and pulse-control constraints including the terminal residual photon number \(N_{\rm res}\).

For the numerical comparisons below, we use the default dissipative point \(\kappa_g=0.64\) and \(\kappa_e=0.04\), giving linewidth ratio \(r_\kappa=\kappa_g/\kappa_e=16\). The dispersive baseline uses the same maximum external linewidth, $\kappa_{\rm disp}= \max(\kappa_g,\kappa_e)=0.64$, and the calibrated half shift \(\chi=0.15\). Unless stated otherwise, we use $\Delta=0$, $N_{\max}=8$, and $t_f=8$, with all drives normalized to satisfy $\max_{s,t}|\alpha_s(t)|^2\le N_{\max}$. 
Details of parameter calibration, pulse normalization, scan protocols, and additional resource checks are given in the Supplemental Material~\cite{SupplementalMaterial}.

\begin{figure}[t]
    \centering
    \includegraphics[width=0.48\textwidth]{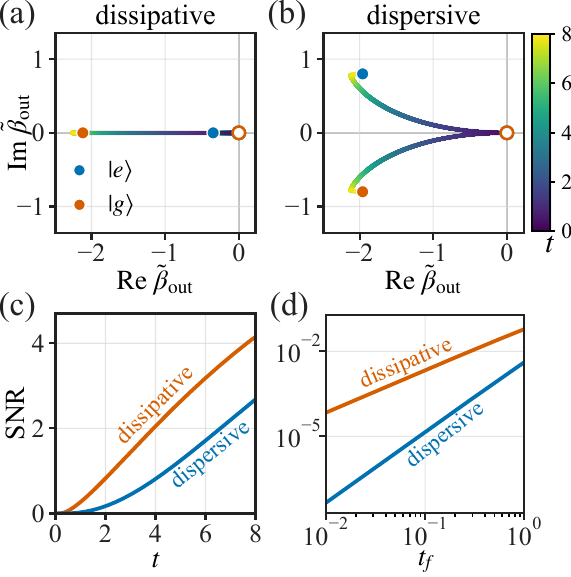}
    \caption{
        Dynamical output response and SNR buildup.
        (a),(b) Background-subtracted conditional output trajectories for the dissipative channel and the dispersive baseline, with
        $\tilde\beta_{\rm out}^{(s)}(t)=\beta_{\rm out}^{(s)}(t)-\beta_{\rm in}(t)$.
        The color scale indicates time; open and filled markers denote $t=0$ and $t=t_f$.
        (c) Time-resolved SNR for the smooth drive.
        (d) Short-time SNR buildup under constant input driving.
        }
    \label{fig:transients}
    \end{figure}

We first use the numerical trajectories to verify the pointer dynamics derived above.
The trajectories in Figs.~\ref{fig:transients}(a,b) show the conditional output records in the complex output-amplitude plane. The dissipative channel separates the two states mainly through the linewidth-dependent amplitude response, whereas the dispersive baseline separates them mainly through conditional phase rotation.
The corresponding SNR buildup under the smooth pulse is shown in Fig.~\ref{fig:transients}(c), while the constant-drive data in Fig.~\ref{fig:transients}(d) approach the asymptotic scalings ${\rm SNR}\propto t_f^{3/2}$ and ${\rm SNR}\propto t_f^{5/2}$.
Because the SNR integrates $|\Delta\beta_{\rm out}(t)|^2$ over the full record, the early $O(t)$ dissipative response remains important even when the dispersive separation grows at later times.

\begin{figure}[t]
    \centering
    \includegraphics[width=0.46\textwidth]{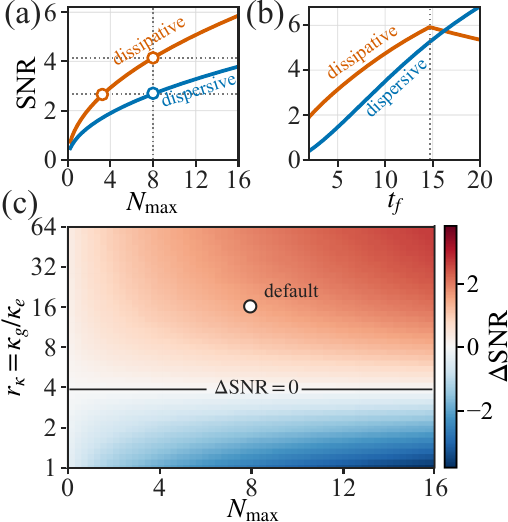}
    \caption{
    Finite-resource SNR scans.
    (a) SNR versus the photon-number ceiling \(N_{\max}\).
    (b) SNR versus measurement readout duration \(t_f\); the vertical dotted line marks the active photon-number constraint switch.
    (c) SNR difference \(\Delta{\rm SNR}\equiv{\rm SNR}_{\rm diss}-{\rm SNR}_{\rm disp}\) over the \((r_\kappa,N_{\max})\) plane; positive values indicate a dissipative advantage, and the marker denotes the default point.
    }
    \label{fig:resource_scans}
    \end{figure}

We next test how the same separation mechanism behaves as the available readout resources are varied.
The scan over photon-number ceiling in Fig.~\ref{fig:resource_scans}(a) shows that the dissipative channel gives a larger SNR throughout the range considered.
At the default value \(N_{\max}=8\), the dissipative channel reaches \({\rm SNR}\simeq4.15\), compared with \({\rm SNR}\simeq2.68\) for the dispersive baseline.
Equivalently, the dispersive default SNR is reached by the dissipative channel at a substantially smaller photon-number budget.
The readout-duration scan in Fig.~\ref{fig:resource_scans}(b) shows that the advantage is a finite-time effect. 
The dissipative channel has a larger SNR over the short- and intermediate-time window, whereas the dispersive response eventually overtakes it at longer times.
The change of slope near the dotted line is caused by the photon-number normalization: the conditional response that saturates \(N_{\max}\) switches as \(t_f\) is varied.
Fig.~\ref{fig:resource_scans}(c) maps \(\Delta{\rm SNR}\equiv{\rm SNR}_{\rm diss}-{\rm SNR}_{\rm disp}\)
over the \((r_\kappa,N_{\max})\) plane. 
At fixed maximum linewidth, increasing \(r_\kappa\) increases the contrast between the two conditional linewidths, and hence between the corresponding radiative amplitudes.
The contour \(\Delta{\rm SNR}=0\) therefore marks the approximate linewidth contrast required for the dissipative channel to outperform the dispersive baseline.

Finally, we benchmark the dissipative channel against pulse-optimized dispersive controls within the same finite-dimensional piecewise-complex control basis. Each control pulse is discretized into $18$ complex-amplitude segments, and each segment is evaluated with $14$ substeps in the linear response calculation. All candidate pulses are evaluated at the same common detuning $\Delta_\ast=0$, under the same resource budget $N_{\max}=8$, $\kappa_{\max}=0.64$, and $t_f=8$, and are renormalized to satisfy $\max_{s,t}|\alpha_s(t)|^2\le N_{\max}$. The main optimizer is a derivative-free, multi-start random local search with an annealed acceptance criterion; a gradient-based L-BFGS-B optimizer~\cite{ZhuLBFGSB1997} is used as an independent cross-check.

We compare three control objectives: pure SNR optimization, active depletion, and a soft residual-photon penalty. Pure SNR optimization imposes no additional input-power constraint beyond the common photon-number normalization. Active depletion keeps the early part of the pulse freely controlled and uses the last six control segments to solve a minimum-norm tail pulse enforcing the hard terminal constraint $\alpha_e(t_f)=\alpha_g(t_f)=0$. The residual-penalty objective optimizes
\begin{equation}
\begin{aligned}
\mathcal J
&=
\mathrm{SNR}
-
\lambda_{\rm res}\sqrt{N_{\rm res}},
\\
N_{\rm res}
&=
\max_s|\alpha_s(t_f)|^2,
\qquad
\lambda_{\rm res}=0.5.
\label{eq:residual-penalty-objective}
\end{aligned}
\end{equation}
These three objectives respectively test maximum output distinguishability, terminal cavity depletion, and a proxy for post-readout reset cost.

\begin{figure}[t]
    \centering
    \includegraphics[width=0.48\textwidth]{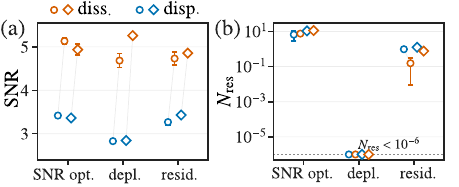}
    \caption{
        Pulse-optimized SNR and residual photons.
        (a) SNR for the dissipative channel and the pulse-optimized dispersive baseline under three objectives: SNR optimization, active depletion, and residual-photon penalty.
        (b) Terminal residual photon number $N_{\rm res}=\max_s|\alpha_s(t_f)|^2$ for the same tests.
        Circles and error bars show the mean and standard deviation over derivative-free multi-start searches; diamonds show the L-BFGS-B cross-check.
        Gray lines pair dissipative and dispersive results for the same objective and optimizer.
        The dashed line marks the plotting floor $N_{\rm res}=10^{-6}$.
        }
    \label{fig:optimal_control}
\end{figure}

Figure~\ref{fig:optimal_control}(a) shows that the dissipative channel remains above the pulse-optimized dispersive baseline for all three objectives and for both optimization strategies. The active-depletion and residual-penalty cases test whether the SNR gain survives when terminal cavity population is also controlled. As shown in Fig.~\ref{fig:optimal_control}(b), both mechanisms can be driven into a low-residual-photon regime, while the dissipative channel still retains the larger SNR. The agreement between the derivative-free search and the L-BFGS-B cross-check indicates that this trend is stable within the chosen finite-dimensional control basis.

\textit{Discussion and conclusions.}---We have formulated a linewidth-encoded dissipative readout channel in which the qubit state controls the radiative linewidth of the readout mode rather than its resonance frequency. At the effective Markov level, this corresponds to a state-dependent jump operator \(L=\sqrt{\kappa(\sigma_z)}a\). The purely dissipative operating point aligns the conditional detunings while keeping different external linewidths,
\(\kappa_e\neq\kappa_g\) and \(\Delta_e=\Delta_g\), so the measured information is carried by the system-output coupling itself. This distinguishes the protocol from Purcell filtering, which engineers the environmental spectrum to suppress qubit decay near \(\omega_q\)~\cite{ref26,ref27,ref28,GoviaClerkSqueezing2017}, and from longitudinal readout~\cite{ref29,ref30,ref31,IkonenMultichannel2019,BottcherLongitudinal2022,HarptLongitudinalReadout2025,XuCatLongitudinal2026}, where the pointer is generated by a Hamiltonian displacement force.

The results established here demonstrate a finite-time SNR advantage across the finite-resource comparisons considered above. The present analysis is an effective-channel theory. A concrete implementation must still verify that the engineered state-dependent linewidth dominates over parasitic transverse coupling, residual state-dependent frequency shifts, Kerr nonlinearities, and Purcell-like relaxation channels. In the convention used here, the excited state is assigned to the weakly radiating branch, \(\kappa_e<\kappa_g\), but protection against qubit relaxation must ultimately come from suppressing parasitic transverse channels rather than from the linewidth assignment alone.

Recent bath-mediated circuit-QED experiments have observed qubit-state-dependent cavity linewidths~\cite{WangNonreciprocity2024}, indicating that conditional radiative damping is experimentally accessible. The Supplemental Material also outlines a possible auxiliary-mode route in which a qubit-dependent auxiliary susceptibility is converted into an effective state-dependent readout linewidth, and includes a comparison with longitudinal-force readout and implementation-level nonideality estimates~\cite{SupplementalMaterial}.

More broadly, these results suggest treating the readout-port coupling itself as a design element of quantum measurement. The engineered jump amplitude carries the signal directly, making the rate at which distinguishability enters the output record a tunable property of the measurement channel. Natural extensions include deriving \(\kappa(\sigma_z)\) from full circuit models, optimizing pulses in larger control spaces, engineering multi-qubit jump amplitudes for parity measurement, and combining linewidth-encoded readout with Purcell-filter protection near \(\omega_q\).

\textit{Acknowledgments.}---Y.-H.C. was supported by the National Natural Science Foundation of China under Grants No. 12304390 and No. 12574386, the National Postdoctoral Overseas Talent Recruitment Program of China, the Fujian 100 Talents Program, and the Fujian Minjiang Scholar Program.
Y.X. was supported by the National Natural Science Foundation of China under Grant No. 62471143 and the Key Program of National Natural Science Foundation of Fujian Province under Grant No. 2024J02008.
F.N. was supported in part by the Japan Science and Technology Agency (JST) via the CREST Quantum Frontiers program Grant No. JPMJCR24I2, the Quantum Leap Flagship Program (Q-LEAP), the Moonshot R\&D Grant No. JPMJMS2061, and the Office of Naval Research (ONR) Global (via Grant No. N62909-23-1-2074).

\bibliography{reference}
\end{document}